\documentclass{article}
\usepackage{amssymb,amsmath}
\newcommand{\PSbox}[3]{\mbox{\rule{0in}{#3}\hspace{#2}\includegraphics{#1}}}

\newtheorem{thm}{Theorem}[section]
\newtheorem{lemma}[thm]{Lemma}

\newtheorem{prop}[thm]{Proposition}

\newcommand{\R}{{\mathbb R}}
\newcommand{\E}{{\mathbb E}}

\newcommand{\Z}{{\mathbb Z}}
\renewcommand{\H}{{\mathbb H}}

\newcommand{\eps}{\epsilon}
\newcommand{\vep}{\varepsilon}

\renewcommand{\Re}{\mbox{Re}}
\newenvironment{proof}{{\bf Proof. }}{\hfill$\square$ \vskip .5cm}
\newcommand{\barh}{{h_0}}

\begin{document}
\title{Dominos and the Gaussian free field.}
\author{Richard Kenyon
\thanks{Laboratoire de Topologie et Dynamique, 
UMR 8628 du CNRS, B\^at 425, Universit\'e Paris-Sud,
91405 Orsay, France.}}
\date{}
\maketitle
\abstract{We define a scaling limit of the height function on
the domino tiling model (dimer model) on simply-connected regions in $\Z^2$ 
and show that it is the ``massless free field'', a Gaussian process
with independent coefficients when expanded in the eigenbasis
of the Laplacian.
}

\begin{center}
{\bf R\'esum\'e}\end{center}
{\small On d\'efinit une ``limite d'\'echelle" pour la fonction
d'hauteur dans le mod\`ele des dim\`eres dans $\Z^2$.
Nous montrons que la limite est un ``champ libre gaussien'',
un processus stochastique gaussien dont les coefficients,
dans la base des fonctions propres du laplacien, sont ind\'ependentes.}
\medskip

\noindent keywords: dimer model, domino tiling, Gaussian free field.

\section{Introduction}
A {\bf domino tiling} of a polyomino $P$ in $\Z^2$ is a tiling of 
$P$ with $2\times 1$ and $1\times 2$ rectangles. For a polyomino $P$
let $\mu=\mu(P)$ denote the uniform measure on the set of all domino tilings of $P$.

Let $U\subset\R^2$ be a Jordan domain with smooth boundary.
We study uniform random domino tilings of polyominos $P_\eps$ in $\eps\Z^2$
which approximate $U$ (and using dominos which are $2\eps\times \eps$ and 
$\eps\times 2\eps$ rectangles). 

\begin{figure}[htbp]
\PSbox{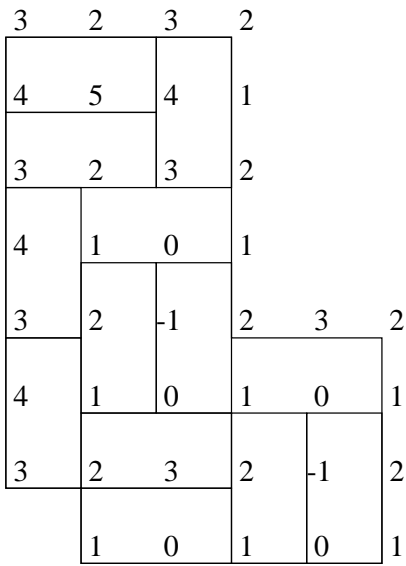}{2in}{3in}
\caption{\label{1} Height function of a domino tiling.}
\end{figure}

A domino tiling of a polyomino $P_\eps$ in $\eps\Z^2$
can be thought of as a random map from $\eps\Z^2\cap P_\eps$ to $\Z$ in the following
way.
Let $V_\eps=\eps\Z^2\cap P_\eps$ be the set of lattice points in the polyomino
$P_\eps$. 
Let $h\colon V_\eps\to\Z$ be a function which has the property that around every
lattice square of $P_\eps$ the four values of $h$ are $4$ consecutive integers
$h_0,h_0+1,h_0+2,h_0+3$, with the values on any two adjacent boundary vertices of $P_\eps$
differing by $1$.
The set of such functions $h$ (up to additive constants and a global sign change)
is in bijection with the set of domino tilings of $P_\eps$: dominos cross exactly
those edges whose $h$-difference is $3$. The function $h$ associated to a tiling is
called its {\bf height function} \cite{Thu}. See Figure \ref{1}.
Note that the height function takes values in $\Z$, not in $\eps\Z$.

Our aim is to prove that in the limit as $\eps\to0$ the height function on a random
tiling of $P_\eps$ tends to a random (generalized) function which has a 
succinct description in terms of the eigenbasis of the Laplacian operator on $U$.

\begin{thm}\label{main}
Let $U$ be a Jordan domain with smooth boundary in $\R^2$. For each $\eps>0$
sufficiently small let $P_\eps$ be a Temperleyan polyomino approximating
$U$ as described below. Let $h_\eps$ be the height of a random
domino tiling of $P_\eps$ and $\bar h_\eps$ be its mean value. 
Then as $\eps$ tends to $0$, $h_\eps-\bar h_\eps$ tends
weakly in distribution to $4/\sqrt{\pi}$ times the ``massless free field'' $F$ on $U$,
in the sense that for any smooth function $\phi$ on $U$,
the random variable $\sum_{x\in V_\eps}\phi(x)(h_\eps(x)-\overline{h_\eps(x)})$ 
tends in distribution to $\frac{4}{\sqrt{\pi}}\int_U\phi F$. 
\end{thm}

For the definition of Temperleyan polyominos see below.
The {\bf massless free field} $F$ on $U$ is a random variable
taking values in the space of distributions
\footnote{Henceforth we will refer to these objects as
``generalized functions'' to avoid confusion.}
which are continuous linear functionals
on the space of $C^1$ functions on $U$ (with a $C^1$-norm).
For background on the massless free field see \cite{Sim}.
It can be defined as follows:
let $\{f_i\}_{i\geq1}$ be an $L^2$-orthonormal
eigenbasis for the Laplacian $\Delta=\frac{\partial^2}{\partial x^2}+
\frac{\partial^2}{\partial y^2}$ on $U$ with
Dirichlet boundary conditions (that is, $f_i\equiv 0$ on $\partial U$).
Let $\lambda_i$ be the eigenvalue of $f_i$. 
Then 
\begin{equation}\label{series}
F=\sum_{i\geq1} \frac{c_i f_i}{(-\lambda_i)^{1/2}},
\end{equation}
where the $c_i$ are i.i.d. \!\!Gaussian random variables of mean $0$ and variance $1$.
Here this expression is interpreted as the generalized function $F$
satisfying, for any $C^1$ function $\phi$,
$$\int_U \phi F=\sum_{i\geq1} \frac{c_i}{(-\lambda_i)^{1/2}}\int_U \phi f_i,$$
series which converges almost surely.
The expression (\ref{series}) does not define a function since the series 
diverges almost everywhere.

\noindent{\bf Remarks.}

\noindent{\bf 1.} The above theorem describes the limiting value of $h_\eps-\bar h_\eps$.
The limiting average value $\bar h=\lim \bar h_\eps$ was computed in \cite{Kenyon.ci}:
it is a harmonic function whose boundary values are given by $\frac2\pi$ times
the angle of turning of the boundary tangent counterclockwise from a fixed basepoint.
(Regarding the choice of basepoint, see the definition of ``Temperleyan" polyomino 
below.)

\noindent{\bf 2.} Theorem \ref{main} has a well-known one-dimensional analog: 
Let $X$ be the sets of random maps 
$h$ from ${0,\frac1{n},\frac{2}{n}\dots,1}$ to $\Z$ satisfying $h(0)=h(1)=0$ and
$|h(\frac{i+1}n)-h(\frac{i}n)|=1$. A random element of $X$, when divided by $\sqrt{n}$,
converges to a random function known as the ``Brownian bridge'' \cite{Levy}. 
In the eigenbasis of the one-dimensional Laplacian $\frac{\partial^2}{\partial x^2}$
the coefficients of the Brownian bridge are again independent Gaussians. 
One difference between the one-dimensional case and Theorem \ref{main}, however, is 
that the height function $h$ of Theorem \ref{main} is {\it unnormalized}.
It is therefore all the more surprising that the integer-valued function $h$ 
of Theorem \ref{main} converges to a continuous-valued object.

\noindent{\bf 3.} An important open problem is to compute the
distribution of the height function on a non-simply connected domain, even
an annulus.
In particular for an annulus the distribution of the height difference
between the two boundary components (in the limit $\eps\to0$)
is unknown although it was shown in \cite{Kenyon.ci}
to depend only on the conformal modulus of the annulus.  

\noindent{\bf 4.} Temperley \cite{Temp} gave a bijection between the {\bf uniform
spanning tree process} on subgraphs of $\Z^2$ and domino tilings. The function $h$ of 
Theorem \ref{main} corresponds under this bijection to the ``winding number''
of the branches of a spanning tree \cite{Kenyon.ci}, as first conjectured by I. Benjamini. 
As it is an open question to show that a scaling limit exists
for the uniform spanning tree process \cite{ABNW,Sch}, one might hope that
the reconstruction of the tree from its winding numbers, which is possible for
$\eps>0$, also works in the limit $\eps=0$. So far this remains an open problem.

\noindent{\bf 5.} The result of Theorem \ref{main} depends strongly on the
choice of boundary conditions for the approximating polyominos $P_\eps$.
For even slight generalizations of these boundary conditions our methods
will not work: see \cite{Kenyon.ci} for a discussion of this issue.

\noindent{\bf 6.} When the region $U$ is a rectangle
$U=[0,a]\times[0,b]$, the orthonormal
eigenvectors of $\Delta$ with Dirichlet boundary conditions are 
$\frac{4}{ab}\sin\frac{\pi j x}a\sin\frac{\pi k y}b$, where $j,k$ are positive
integers. So in this case the massless free field has independent
Fourier coefficients.

\noindent{\bf 7.} Most of the work to prove Theorem \ref{main}
was done in \cite{Kenyon.ci}, where we proved Proposition \ref{moms},
below.
\medskip

If we consider the massless free field $F$ to be a continuous linear functional
on the space of smooth $2$-forms on $U$ (rather than on the space of 
smooth functions on $U$)
then $F$ is {\bf conformally invariant}, in the following sense.
\begin{prop}\label{confinv}
Let $\omega$ be a smooth $2$-form on $U$ and let $f\colon V\to U$
be a conformal bijection. Let $F_U,F_V$ be the massless free fields on $U$ and $V$
respectively. Let $X = \int_U F_U(z)\omega(z)$ and 
$Y= \int_V F_V(z) f^*\omega(z),$ where $f^*\omega$ is the pullback of
$\omega$ to $V$. Then the random variables $X$ and $Y$ are equal in distribution.
\end{prop}

For the proof see section \ref{propproof}.
\section{Background and preliminaries}
\subsection{Temperleyan polyominos and approximation}\label{temper}
Define the $(i,j)$-lattice square in $\Z^2$ to be the lattice square 
whose lower left corner is $(i,j)$. A lattice square is said to be {\it even}
if the coordinates of its lower left corner are even.
A {\bf polyomino} 
is a union of lattice squares which is bounded by a simple closed lattice curve.
A polyomino is {\bf even} if all of its
{\it corner squares} are even, where
by corner squares we mean those lattice squares adjacent to the corners and containing
the interior angle bisector at the corner.
In particular note that an edge of an even polyomino
$P'$ has odd length if its two extremities
are both concave or both convex corners; 
if the extremities consist of one concave and one convex
corner the edge length is even.
Let $P$ be a polyomino
obtained from an even polyomino $P'$ by removing one lattice square $b$
adjacent to its boundary, where $b$ is of the same parity as the corners of $P'$. 
Such a polyomino is called {\bf Temperleyan}, and the removed square 
$b$ is called its {\bf root}. In Figure \ref{1}, the polyomino is Temperleyan
with root the lower left (removed) square. 

All Temperleyan polyominos have domino tilings \cite[section 7]{Kenyon.ci}. 
The term {\it Temperleyan} comes from the bijection due to Temperley between the
set of spanning trees of a rectangle in $\Z^2$ and  the set of domino tilings of 
a rectangular region with a corner removed \cite{Temp}. This bijection
was generalized in \cite{BP} and further in \cite{KPW}.

Let $U$ be a smooth Jordan domain with a marked point $b\in\partial U$.
For each $\eps>0$ let $P_\eps$ be a Temperleyan polyomino in $\eps\Z^2$
approximating $U$ as follows. The boundary of $P_\eps$ lies within $O(\eps)$ of
$\partial U$, and the counterclockwise
boundary path of $P_\eps$ points locally
into the same half-space as the (directed) tangent to $\partial U$ which
it is near. Furthermore the root $b_\eps$ of $P_\eps$ should be within $O(\eps)$ of $b$.

\subsection{Green's functions}
Let $U$ be a Jordan domain with basepoint $b\in\partial U$. 
The Green's function with Dirichlet boundary conditions, or simply Dirichlet
Green's function, $g_D(z_1,z_2)$, is defined to be the 
unique function (of $z_2$)
satisfying $\Delta g_D(z_1,z_2)=\delta_{z_1}(z_2)$ (the Dirac delta), 
and which is zero when $z_2\in\partial U$,
where the Laplacian is with respect to the second variable.
This function is well-defined and when $z_2$ is near $z_1$ has the 
form $g_D(z_1,z_2)=\frac1{2\pi}\log|z_2-z_1|+O(1)$.

The Dirichlet Green's function has the following simple expression in the 
basis of eigenfunctions of the Laplacian on $U$:
\begin{lemma}
$$g_D(z_1,z_2) = \sum_{i\geq 1}\frac{f_i(z_1)f_i(z_2)}{\lambda_i}.$$
\end{lemma}
\begin{proof}
Since the eigenbasis $\{f_i\}$ of $\Delta$ is an orthonormal basis for $L^2(U)$, it suffices
to show that for each $i$, $\langle f_i(z_2),g_D(z_1,z_2)\rangle =  \frac{f_i(z_1)}{\lambda_i}$.
But  
\begin{eqnarray*}
\langle f_i(z_2),g_D(z_1,z_2)\rangle &=&
\frac1{\lambda_i}\langle \lambda_i f_i(z_2),g_D(z_1,z_2)\rangle\\
&=& \frac1{\lambda_i}\langle \Delta f_i(z_2),g_D(z_1,z_2)\rangle\\
&=& \frac1{\lambda_i}\langle f_i(z_2),\Delta g_D(z_1,z_2)\rangle\\
&=& \frac1{\lambda_i}\langle f_i(z_2),\delta_{z_1}(z_2)\rangle\\
&=& \frac1{\lambda_i}f_i(z_1).
\end{eqnarray*}
\end{proof}

We will also need to define the Neumann Green's function $g_N$.
On a bounded domain,
the Green's function with Neumann boundary conditions does not exist. However we
can define (see below)
the function which is the ``difference of two Neumann Green's functions'':
for points $z_1,z_1'\in U$
define the difference of Neumann Green's
functions $g_N(z_1,z_2)-g_N(z_1',z_2)$ to be the function of $z_2$
satisfying 
$\Delta(g_N(z_1,z_2)-g_N(z_1',z_2))=\delta_{z_1}(z_2)-\delta_{z_1'}(z_2)$, where the Laplacian
is with respect to the second variable, and which satisfies
$\frac{\partial}{\partial \hat n}(g_N(z_1,z_2)-g_N(z_1',z_2))=0,$ that is, 
the derivative normal to the boundary
with respect to the second variable is zero.
This function is well-defined up to an additive constant,
and we set the constant so that $g(z_1,b)-g(z_1',b)=0$. 

Near $z_1$ we have $g_N(z_1,z_2)-g_N(z_1',z_2)=\frac1{2\pi}\log|z_2-z_1|+c_1(z_1,z_1')+O(z_2-z_1).$
Let $\hat g_N(z_1,z_2)-\hat g_N(z_1',z_2)$ be the harmonic conjugate 
(with respect to the second variable) of $g_N(z_1,z_2)-g_N(z_1',z_2)$.
This function is multiply-valued, increasing by $1$ when $z_2$ turns counterclockwise
around $z_1$ and by $-1$ when $z_2$ turns counterclockwise 
around $z_1'$.
The function $\tilde g_N(z_1,z_2)-\tilde g_N(z_1',z_2):=g_N(z_1,z_2)-g_N(z_1',z_2)+
i(\hat g_N(z_1,z_2)-\hat g_N(z_1',z_2))$ 
is analytic in $z_2$ (except at $z_1$ and $z_1'$)
and is the {\it analytic Neumann Green's function}. It is also multiply-valued.

We can define the exterior derivative of $\tilde g_N$ with respect to the first variable
as $d\tilde g_N(z_1,z_2)= F_0(z_1,z_2)dx_1 + iF_1(x,z_2)dy_1$, 
where $z_1=x_1+iy_1$ and $F_0,F_1$ are defined
by taking limits (for $\delta$ real)
$F_0(z_1,z_2)=\lim_{\delta\to0}\frac{\tilde g_N(z_1+\delta,z_2)-\tilde g_N(z_1,z_2)}{\delta}$ and 
$iF_1(z_1,z_2)=\lim_{\delta\to0}\frac{\tilde g_N(z_1+i\delta,z_2)-\tilde g_N(z_1,z_2)}{\delta}$.
These functions $F_0,F_1$ are well-defined (single-valued) and vanish at $z_2=b$.

As examples of these functions, on the upper half-plane $\H$ with $b=\infty$ we have
\begin{equation}\label{dgreen}
g_D(z_1,z_2)=\frac1{2\pi}\log\left|\frac{z_2-z_1}{z_2-\bar z_1}\right|,
\end{equation}
and 
\begin{equation}\label{an}
\tilde g_N(z_1,z_2)-\tilde g_N(z_1',z_2) = 
\frac1{2\pi}\log\frac{(z_2-z_1)(z_2-\bar z_1)}{(z_2-z_1')(z_2-\overline{z_1'})}.
\end{equation}
Note that when $z_2\in\R$ the imaginary part of (\ref{an}) is constant
(in fact vanishes); this implies
that the real part has Neumann boundary conditions.
For a more general Jordan domain $V$, let $f$ be a Riemann map from $V$ to the upper
half-plane sending $b$ (the base point of $V$)
to $\infty$. Then the Dirichlet Green's function on 
$V$ is $g^V_D(z_1,z_2)=g_D^{\H}(f(z_1),f(z_2))$, and the analytic Neumann Green's
function is defined similarly $\tilde g_N^V(z_1,z_2)-\tilde g^V_N(z_1',z_2) = 
\tilde g_N^{\H}(f(z_1),f(z_2))-\tilde g^{\H}_N(f(z_1'),f(z_2))$.
One can in fact take this to be the definition of the Green's functions on $V$.

\subsection{Moment formula}
For a region $U$ with basepoint $b\in\partial U$,
define the functions $F_+(z_1,z_2)$ and $F_-(z_1,z_2)$ by
\begin{equation}\label{Fpmdef}
-4d\tilde g_N(z_1,z_2) = F_+(z_1,z_2)dz_1 + F_-(z_1,z_2)d\overline{z_1},
\end{equation}
where $d$ is exterior differentiation with respect to the first variable.
Then $F_+(z_1,b)=0=F_-(z_1,b)$
(in terms of the functions $F_0,F_1$ of the previous section we have
$F_\pm=-2(F_0\pm F_1)$).

Let $\barh(x) = h(x)-\bar h(x)$.

\begin{prop}[\cite{Kenyon.ci}]\label{moms} Under the hypotheses of Theorem \ref{main},
let $z_1,\dots,z_k$ be distinct points of $U$, and $\gamma_1,\dots,\gamma_k$
disjoint paths running from the boundary of $U$ to $z_1,\dots,z_k$ respectively. 
Let $h(z_1)$ denote the height of a point of $P_\eps$ lying within
$O(\eps)$ of $z_1$.
Then
\begin{equation}\label{mom}\lim_{\eps\to0}\E(\barh(z_1)\cdots \barh(z_k))= 
\sum_{\vep_1,\dots,\vep_k\in\{\pm1\}}\vep_1\cdots \vep_k\int_{\gamma_1}\cdots\int_{\gamma_{k}}\det_{i,j\in[1,k]}\Bigl(
F_{\vep_i,\vep_j}(z_i,z_j)\Bigr)dz_1^{(\vep_1)}\cdots
dz_k^{(\vep_k)},
\end{equation}
where $dz_j^{(1)}=dz_j$ and $dz_j^{(-1)}=d\overline{z_j}$, and
$$F_{\vep_i,\vep_j}(z_i,z_j)=\left\{\begin{array}{ll}
0&\mbox{\rm if }i=j\\
F_+(z_i,z_j)&\mbox{\rm if }(\vep_i,\vep_j)=(1,1)\\
F_-{(z_i,z_j)}&\mbox{\rm if }(\vep_i,\vep_j)=(-1,1)\\
\overline{F_-{(z_i,z_j)}}&\mbox{\rm if }(\vep_i,\vep_j)=(1,-1)\\
\overline{F_+{(z_i,z_j)}}&\mbox{\rm if }(\vep_i,\vep_j)=(-1,-1).\end{array}\right.$$
\end{prop}

\section{Proof of Theorem \protect\ref{main}}
When $U$ is the upper half plane with basepoint at $\infty$, the derivative of the 
analytic Neumann Green's function is (see (\ref{an}))
$$d\tilde g_N(z_1,z_2) = 
\frac{dz_1}{2\pi(z_1-z_2)}+\frac{d\overline{z_1}}{2\pi(\overline{z_1}-z_2)}.$$
Thus from (\ref{Fpmdef}) we have $F_+(z_1,z_2)=\frac{2}{\pi(z_2-z_1)}$ and
$F_-(z_1,z_2)=\frac2{\pi(z_2-\overline{z_1})}$.

Let $p,q\in U$. From Proposition \ref{moms} we have $\lim_{\eps\to0}\E(h_0(p)h_0(q))=$
$$=
\int_{\gamma_1,\gamma_2}
\left|\begin{array}{cc}0&F_+(z_1,z_2)\\F_+(z_2,z_1)&0\end{array}\right|dz_1dz_2 -\int_{\gamma_1,\gamma_2}
\left|\begin{array}{cc}0&F_-(z_1,z_2)\\\overline{F_-(z_2,z_1)}&0\end{array}\right|d\overline{z_1}dz_2 -\hskip1in$$
$$\hskip1in-\int_{\gamma_1,\gamma_2}
\left|\begin{array}{cc}0&\overline{F_-(z_1,z_2)}\\F_-(z_2,z_1)&0\end{array}\right|dz_1d\overline{z_2} +
\int_{\gamma_1,\gamma_2}
\left|\begin{array}{cc}0&\overline{F_+(z_1,z_2)}\\\overline{F_+(z_2,z_1)}&0\end{array}\right|d\overline{z_1}d\overline{z_2}.$$
Plugging in for $F_\pm$ gives
$$\lim_{\eps\to0}\E(h_0(p)h_0(q))=
-\frac4{\pi^2}\int_{\gamma_1}\int_{\gamma_2}\frac1{(z_2-z_1)^2}dz_1dz_2 +
\frac4{\pi^2}\int_{\gamma_1}\int_{\gamma_2}\frac1{(z_2-\overline{z_1})^2}
d\overline{z_1}dz_2+
\qquad\qquad$$
$$\qquad\qquad + \frac4{\pi^2}\int_{\gamma_1}\int_{\gamma_2}\frac1{(\overline{z_2}-z_1)^2}
dz_1d\overline{z_2}
-\frac4{\pi^2}\int_{\gamma_1}\int_{\gamma_2}\frac1{(\overline{z_2}-\overline{z_1})^2}
d\overline{z_1}d\overline{z_2} $$
\begin{equation}\label{star}=\frac8{\pi^2}\Re\log\left(\frac{\overline{p}-q}{p-q}\right).
\end{equation}

Note that this is $-\frac{16}{\pi}g_D(p,q)$ where
$g_D$ is the Dirichlet Green's function on $U$ (see (\ref{dgreen})).

Now let $p_1,\dots,p_k$ be distinct points in the upper half plane $U$.
By symmetry of $\barh$, if $k$ is odd the moment
$\E(\barh(p_1)\cdots\barh(p_k))$ is zero. 
We therefore assume $k$ is even.
In (\ref{mom}), the matrix has $ij$ entry $$F_{\vep_i,\vep_j}(z_i,z_j)=
\frac{2}{\pi(z_j^{(\vep_j)}-z_i^{(\vep_i)})}.$$
Such a matrix has a simple determinant:
\begin{lemma}\label{half}
For $k$ even let $M$ be the $k\times k$ matrix $M=(m_{ij})$ 
with $m_{ii}=0$ and $m_{ij}=\frac1{x_j-x_i}$ when $i\neq j$.
Then 
\begin{equation}\label{pairs}
\det(M) = \sum\frac{1}{(x_{\sigma(1)}-x_{\sigma(2)})^2(x_{\sigma(3)}-x_{\sigma(4)})^2
\cdots(x_{\sigma(k-1)}-x_{\sigma(k)})^2},
\end{equation}
where the sum is over all $(k-1)!!$
possible pairings $\{\{\sigma(1),\sigma(2)\},\ldots,\{\sigma(k-1),\sigma(k)\}\}$ of $\{1,\dots,k\}$.
\end{lemma}
This lemma also appears in \cite{ID}.

\begin{proof} The proof is by induction on $k$. The formula clearly holds when $k=2$.
For $k>2$, the determinant is a rational function of $x_1$ with a double pole at $x_1=x_2$;
we can write 
$$\det(M)=\frac{c_{-2}}{(x_1-x_2)^2} + \frac{c_{-1}}{(x_1-x_2)} + c_0 + O(x_1-x_2).$$
The coefficient $c_{-1}$ is zero since
the determinant is even under the exchange of $x_1$ and $x_2$ (exchange 
the first two rows and exchange the first two columns).
The coefficient $c_{-2}$ is the determinant of $M_{12}$, the matrix obtained from
$M$ by deleting the first two rows and columns. 
Therefore the right and left-hand sides of (\ref{pairs})
both represent rational
functions (in each variable)
with the same poles and residues; hence they differ by a constant.
This constant is zero by homogeneity.
\end{proof} 

Combining the lemma with Proposition \ref{moms} gives the following.
\begin{prop}
Let $U$ be a Jordan domain with smooth boundary. Let $p_1,\ldots,p_k\in U$ 
(with $k$ even) be distinct points. We have
$$\lim_{\eps\to0}\E(\barh(p_1)\cdots\barh(p_k))=\left(-\frac{16}{\pi}\right)^{k/2}\sum_{\mbox{pairings } \sigma} g_D(p_{\sigma(1)},p_{\sigma(2)})
\cdots g_D(p_{\sigma(k-1)},p_{\sigma(k)}).$$
\end{prop}

\begin{proof}
When $U$ is the upper half plane this follows by combining 
Proposition \ref{moms} with Lemma \ref{half} and the calculation (\ref{star}).
For arbitrary $U$, equation (\ref{star}) shows that
$\E(\barh(p_1)\barh(p_2))=-\frac{16}{\pi}g^U_D(p_1,p_2)$ (where $g_D^U$ is the 
Dirichlet Green's function on $U$) by conformal invariance
of the height moments and of $g_D$. This completes the proof. 
\end{proof}

The proof of Theorem \ref{main} is completed as follows.
Let $f_{n_1},\dots,f_{n_k}$ be (not necessarily distinct)
eigenvectors of $\Delta$ with Dirichlet boundary conditions.
Let $C^{(\eps)}_{n_j}$ be the real-valued random variable $C^{(\eps)}_{n_j}=
\eps^2\sum_{x\in V_\eps}h_0(x)f_{n_j}(x)$, where the sum is over the vertices $V_\eps$
of $P_\eps$, and $f_{n_j}(x)$ is $f_{n_j}$ evaluated
at the vertex $x$.  We have

\vskip-.6cm
\begin{eqnarray*} 
\lefteqn{\lim_{\eps\to0}\E(C_{n_1}^{(\eps)}\cdots C^{(\eps)}_{n_k})=}\\
&=& \lim_{\eps\to0}\E\left(\sum_{x_1\in V_\eps}\eps^2h_0(x_1)f_{n_1}(x_1)\cdots
\sum_{x_k\in V_\eps}\eps^2h_0(x_k)f_{n_k}(x_k)\right)\\
&=& \lim_{\eps\to0}\sum_{x_1\in V_\eps}\cdots\sum_{x_k\in V_\eps}
\eps^2f_{n_1}(x_1)\cdots \eps^2f_{n_k}(x_k)\E\left(h_0(x_1)\cdots h_0(x_k)\right)\\
&=&\left(-\frac{16}{\pi}\right)^{k/2}\int_U\cdots\int_Uf_{n_1}(x_1)\cdots f_{n_k}(x_k)
\sum_{\sigma} g_D(x_{\sigma(1)},x_{\sigma(2)}) \cdots g_D(x_{\sigma(k-1)},x_{\sigma(k)})\\
&=&\left(-\frac{16}{\pi}\right)^{k/2}\sum_{\sigma}\int_U\cdots\int_Uf_{n_1}(x_1)\cdots 
f_{n_k}(x_k)\sum_{m_1,\dots,m_{k/2}}\frac{f_{m_1}(x_{\sigma(1)})
f_{m_1}(x_{\sigma(2)})}{\lambda_{m_1}}\cdots \frac{f_{m_{k/2}}(x_{\sigma(k-1)})
f_{m_{k/2}}(x_{\sigma(k)})}{\lambda_{m_{k/2}}}\\
&=&\left(\frac{16}{\pi}\right)^{k/2}\sum_{\sigma}
\frac{\delta_{\sigma(1),\sigma(2)}}{(-\lambda_{\sigma(1)})}\cdots
\frac{\delta_{\sigma(k-1),\sigma(k)}}{(-\lambda_{\sigma(k-1)})}.
\end{eqnarray*}
By Wick's theorem \cite{Sim}, these are exactly the moments
for a set of independent Gaussians of mean zero and variances $-\frac{16}{\pi\lambda_i}$.
Now to conclude we invoke the following standard probability
lemma :
\begin{lemma}[\cite{Bill}]
A sequence of
(multidimensional) random variables whose moments converge to the moments of
a Gaussian, converges itself to a Gaussian.
\end{lemma}
This completes the proof.

\section{Proof of Proposition \ref{confinv}}\label{propproof}
Since $X$ and $Y$ are Gaussians (each being the sum of Gaussians), 
and have mean $0$, it suffices to compute their variances.
But 
\begin{eqnarray*}
\E(X^2) &=& \int_U\int_U \omega(z_1)\omega(z_2)\E(F(z_1)F(z_2))\\
&=& \int_U\int_U \omega(z_1)\omega(z_2)g_D^U(z_1,z_2)\\
&=& \int_V\int_V f^*\omega(y_1)f^*\omega(y_2)g_D^U(f(y_1),f(y_2))\\
&=& \int_V\int_V f^*\omega(y_1)f^*\omega(y_2)g_D^V(y_1,y_2)\\
&=& \E(Y^2)
\end{eqnarray*}
where we used the conformal invariance of the Green's function, 
$g_D^U(f(y_1),f(y_2))=g_D^V(y_1,y_2)$.
This completes the proof.

\end{document}